\documentclass[twocolumn]{aastex62}

\newcommand{\ikoma}[1]{\textcolor{black}{#1}}
\newcommand{\ikomaf}[1]{\textcolor{black}{#1}}

\newcommand{\kawashima}{\textcolor{black}}
\newcommand{\kawashimat}{\textcolor{black}}
\newcommand{\kawashimar}{\textcolor{black}}
\newcommand{\kawashi}{\textcolor{black}}
\newcommand{\kawa}[1]{{#1}}

\newcommand{\renyu}{\textcolor{black}}

\received{}
\revised{}
\accepted{}
\submitjournal{ApJL}

\shorttitle{Molecular features above hydrocarbon haze in transit}
\shortauthors{Kawashima et al.}

\begin{document}

\title{Detectable molecular features above hydrocarbon haze via transmission spectroscopy with JWST: \\
Case studies of GJ~1214b\kawa{-}, GJ~436b\kawa{-}, HD~97658b\kawa{-}, and Kepler-51b\kawa{-like planets}}

\correspondingauthor{Yui Kawashima}
\email{y.kawashima@sron.nl}

\author[0000-0003-3800-7518]{Yui Kawashima}
\affiliation{SRON Netherlands Institute for Space Research, Sorbonnelaan 2, 3584 CA Utrecht, The Netherlands}
\affiliation{Earth-Life Science Institute, Tokyo Institute of Technology 2-12-1-IE-1 Ookayama, Meguro-ku, Tokyo 152-8550, Japan}
\affiliation{Department of Earth and Planetary Science, Graduate School of Science, The University of Tokyo, 7-3-1 Hongo, Bunkyo-ku, Tokyo 113-0033, Japan}

\author{Renyu Hu}
\affiliation{Jet Propulsion Laboratory, California Institute of Technology, 4800 Oak Grove Drive, Pasadena, CA 91109, USA}
\affiliation{Division of Geological and Planetary Sciences, California Institute of Technology, 1200 East California Boulevard, Pasadena, CA 91125, USA}

\author{Masahiro Ikoma}
\affiliation{Department of Earth and Planetary Science, Graduate School of Science, The University of Tokyo, 7-3-1 Hongo, Bunkyo-ku, Tokyo 113-0033, Japan}
\affiliation{Research Center for the Early Universe (RESCEU), Graduate School of Science, The University of Tokyo, 7-3-1 Hongo, Bunkyo-ku, Tokyo 113-0033, Japan}



\begin{abstract}
Some of the exoplanets so far observed show featureless or flat transmission spectra, possibly indicating the existence of clouds and/or haze in \ikoma{their} atmospheres. 
Thanks to its large aperture size and broad wavelength coverage, \ikoma{JWST} is expected to enable detail\ikoma{ed} investigation of exoplanet atmospheres\ikoma{, which could provide important constraints on the atmospheric composition obscured by clouds/haze}.
Here, we use four warm ($\lesssim 1000$~K) planets suitable for atmospheric characterization via transmission spectroscopy, GJ~1214b, GJ~436b, HD~97658b, and Kepler-51b, as examples to explore molecular absorption features detectable by JWST even in the existence of hydrocarbon haze in the atmospheres. We simulate photochemistry, the growth of hydrocarbon haze particles, and transmission spectra for the atmospheres of these four planets.
\kawa{Among the planetary parameters considered, {super-Earths with hazy, relatively hydrogen-rich atmospheres} are mostly expected to produce detectable molecular absorption features {such as a quite prominent $\mathrm{CH_4}$ feature at 3.3~${\rm \mu}$m}} \kawa{even for the {extreme} case of the {most efficient} production of photochemical haze.}
\renyu{For a planet that has extremely low gravity, such as Kepler-51b, haze particles grow significantly large in the upper atmosphere due to the small sedimentation velocity, resulting in the featureless or flat transmission spectrum in a wide wavelength range.}
\renyu{This investigation shows that the transmission spectra with muted features measured by \kawa{HST} in most cases do not preclude strong features at the longer wavelengths accessible by JWST.}
\end{abstract}



\section{Introduction} 
\ikoma{T}ransmission spectra of close-in \kawashimar{exoplanets} \ikoma{observed} so far \ikoma{are \kawa{somewhat} diverse} \citep{Sing:2016hi}.
There are some studies that explored \ikoma{such} diversity by examining the correlation between the observed amplitude of absorption features and the planetary properties 
\citep{2016ApJ...817L..16S, 2016ApJ...826L..16H, 2017AJ....154..261C}.
\kawashima{They all reached the conclusion that molecular absorption features are less pronounced in transmission spectra for lower equilibrium temperature although other planetary properties may affect such a correlation.}

One explanation for this \ikoma{correlation} is existence of photochemically-produced hydrocarbon haze obscuring the \ikoma{molecular} feature\ikoma{s} in the atmospheres of cooler planets \kawashimar{\citep{2009arXiv0911.0728Z, 2012ApJ...745....3M, 2013ApJ...775...33M, 2015ApJ...815..110M}}, because its primary source $\mathrm{CH_4}$ exists \ikoma{only at such} low temperature\ikoma{s} ($\lesssim 1000$~K) \citep[e.g.,][]{1999ApJ...512..843B}. 

Among \ikoma{stars hosting} currently observable warm ($\lesssim 1000$~K) \ikoma{low-mass} planets, GJ~1214, GJ~436, and HD~97658 are the only three \ikoma{host stars whose} UV \ikoma{emission} spectra have been observed \citep{2016ApJ...820...89F, Youngblood:2016ib, 2016ApJ...824..102L}. 
\ikoma{Also,} high-precision transmission spectra \ikoma{for their planets were} observed with HST \citep[e.g.,][]{2014Natur.505...69K, 2014Natur.505...66K, 2014ApJ...794..155K}.
They all show flat or featureless transmission spectra \ikoma{between 1.1 and 1.7~$\rm \mu m$}, possibly indicating the existence of haze in the atmospheres. Since the knowledge of \ikoma{host-star's} UV spectrum is essential for the modeling of hydrocarbon haze 
\ikoma{\citep[e.g.,][]{2018ApJ...853....7K, Kawashima.Ikoma2019}}, the above three planets serve as \renyu{the most promising targets} to understand the nature of haze in exoplanet atmospheres.
In addition, Kepler-51b is attracting attention as a favorable target for atmospheric characterization via transmission spectroscopy due to its extremely low density and large atmospheric scale height \citep{2014ApJ...783...53M}.
In terms of haze science, this planet is also an interesting target because the sedimentation velocity of the particles in the atmosphere is expected to be quite low due to its low gravity, which allow\ikoma{s} particle growth in the upper atmosphere and \renyu{mute features in the transmission spectrum}.
\kawashimar{Recently, \cite{2019ApJ...873L...1W} investigated the effect of dusty outflows for Kepler-51b and 51d and showed that the existence of small dust \kawa{of a fixed size} at high altitudes could flatten the transmission spectra.}

Currently, the number of exoplanets suitable for atmospheric characterization is still small due to the lack of bright targets and sufficient observational precision. Fortunately, 
TESS \citep{2014SPIE.9143E..20R} 
is expected to detect a great number of transiting exoplanets around nearby stars bright enough for atmospheric characterization. Also, 
JWST \citep{2006SSRv..123..485G} 
will enable high-precision transmission spectroscopy thanks to its large diameter, and also enable the spectroscopy at longer wavelengths than HST with its suite of spectroscopy instruments up to \renyu{12~$\mu$m}\kawa{, with} photometry up to 28.5~$\mu$m {\citep{2014PASP..126.1134B}}.

In this Letter, we use the above four favorable planets, GJ~1214b, GJ~436b, HD~97658b, and Kepler-51b, as examples and explore molecular absorption features detectable by JWST \renyu{in the existence of hydrocarbon haze in their atmospheres}.
Planets similar in size to GJ~1214b and HD~97658b have been shown to be abundant \citep{2018AJ....156..264F}, indicating more planets in this size range will be found by TESS and will be the primary targets for atmospheric characterization by JWST \kawashimar{\citep{2018PASP..130d4401L}}.
\renyu{We describe the models in \S\ref{method}, and show the results in \S\ref{results}.} {We 
\kawashimat{discuss several effects to be examined} in \S\ref{discussion}} and conclude this Letter in \S\ref{conclusions}

\section{Method} \label{method}
We model transmission spectra of an atmosphere with hydrocarbon haze using the photochemical, particle growth, and transmission spectrum models of \cite{2018ApJ...853....7K} as follows:
We first perform photochemical calculations to derive the \renyu{steady-state}, distribution of gaseous species. \renyu{Our reaction rate list is the \kawashima{reduced vesion of} \cite{2012ApJ...761..166H} and we include the reverse reactions in the same way as \cite{2014ApJ...784...63H}.}
Then we assume that the production rate of haze monomers at each altitude as the sum of the photodissociation rates of the major hydrocarbons in our photochemical model, $\mathrm{CH_4}$, $\mathrm{HCN}$, and $\mathrm{C_2H_2}$, as an upper limit for the monomer production rate since \renyu{this approach effectively assumes 100\% conversion efficiency from the photodissociation of these hydrocarbons to haze monomers}.
With this assumption, we derive the \renyu{steady-state} distribution of haze particles by the particle growth calculations.
Finally, we model transmission spectra of the atmospheres with the obtained distributions of haze particles and gaseous species.
For the details of each of the three models, see \cite{2018ApJ...853....7K}.

As for the UV spectra of \ikoma{the host stars} GJ~1214, GJ~436, and HD~97658, we use the data constructed by the MUSCLES Treasury Survey \citep{2016ApJ...820...89F, Youngblood:2016ib, 2016ApJ...824..102L}.
As for Kepler-51 \ikoma{whose} UV spectrum has not been observed\ikomaf{,} 
\kawa{because of its similar properties to the Sun,}
we use \ikoma{the solar spectrum} from \cite{2003AsBio...3..689S}\footnote{http://vpl.astro.washington.edu/spectra/stellar/other\_stars.htm}.

\ikoma{The setting for the atmospheres is as follows.}
We assume \ikoma{that the elemental abudance ratios of the atmospheric gas are} the solar system \ikoma{ones}, which we take from Table~2 of \cite{2003ApJ...591.1220L}.
For the temperature-pressure profile, we use the analytical formula of Eq.~(29) of \cite{2010A&A...520A..27G}.
We choose the values of the parameters, namely, the intrinsic temperature $T_\mathrm{int}$, \kawashima{irradiation} temperature $T_\mathrm{irr}$, averaged opacity in the optical $k_\mathrm{v}$ and that in the infrared $k_\mathrm{th}$, so as to match the temperature-pressure profiles from \cite{2010ApJ...716L..74M} (\ikoma{the version with the} solar \ikoma{metallicity and} efficient day\ikoma{side-}to\ikoma{-}nightside heat redistribution) for GJ~1214b and \ikoma{from} \cite{2010ApJ...720..344L} (their solar\ikoma{-metallicity} version) for GJ~436b. 
As for HD~97658b and Kepler-51b, we adopt the same \kawa{parameter} values 
as the case of GJ~1214b \kawa{except for $T_\mathrm{irr}$, which} 
we calculate 
with Eq.~(1) of \cite{2010A&A...520A..27G}.
%
\kawashima{\kawa{We adopt} the lower-boundary pressure \kawa{of 1000~bar} for photochemical calculation, 
while 10~bar in particle growth and transmission spectrum calculations.}

For the calculations of photochemistry and particle growth, we choose the values of the reference \ikoma{radius} \ikoma{equivalent to 1000~bar} so as to roughly match the observed transit radii 
\kawa{for} a clear solar-composition atmosphere. 
Only for Kepler-51b, we use different values of \ikoma{the} 1000-bar \ikoma{radius} for \ikoma{two cases of the} hazy and clear atmospheres. 
\ikoma{This is} because \ikoma{with} the same 1000-bar \ikoma{radius} as \ikoma{in} the clear atmosphere case\ikoma{, we} would \ikoma{obtain} too large \ikoma{transit radii for the hazy atmosphere to be} consistent with the observed depth due to its extremely low \kawa{gravity}.
When calculating the transmission \ikoma{spectra}, we find the appropriate value of 10-bar radius that \ikoma{minimizes the} reduced $\chi^2$ value \ikoma{in comparison between} the theoretical and observed transit depths. {We consider the observed transmission spectra of \cite{2014Natur.505...69K}, \cite{2014Natur.505...66K}, \cite{2014ApJ...794..155K} for GJ~1214b, GJ~436b, and HD~97658b, respectively, and the observed radius of Kepler-51b from \cite{2014ApJ...783...53M}.} 
For the calculation of theoretical transit depth for each observed data point, we consider the transmission curve of the filter used in the observation\kawashimar{, taking} those data 
from the SVO Filter Profile Service\footnote{http://svo2.cab.inta-csic.es/theory/fps/} \citep{2012ivoa.rept.1015R, Rodrigo2013}.

\ikoma{The} parameters and their values we use are listed in Table~\ref{tab:parameters}.

\begin{deluxetable*}{llll}
\tablecaption{Model parameters and their values used in the simulations \label{tab:parameters}}
\tablehead{
\colhead{Parameter} & \colhead{Description} & \colhead{Value} & \colhead{Reference} \\
}
\startdata
Common parameters \\
$K_{zz}$ & Eddy diffusion coefficient & $1.00 \times 10^7$~$\mathrm{cm^2}$~$\mathrm{s^{-1}}$ & \\
$s_1$ & Monomer radius & $1.00 \times 10^{-3}$~$\mu$m & \\
$\rho_p$ & Particle internal density & $1.00$~g~$\mathrm{cm^{-3}}$ & \\ \hline
GJ~1214b \\
$R_s$ & Host star radius & 0.201~$R_\odot$ & \citet{2013AA...551A..48A} \\
$M_p$ & Planet mass & $6.26$~$M_\oplus$ & \citet{2013AA...551A..48A} \\
$R_{1000}$~$_\mathrm{bar}$ & 1000-bar radius & 2.07~$R_\oplus$ & \\
$a$ & Semi-major axis & 0.0148~AU & \citet{2013AA...551A..48A} \\
$d$ & Distance & 14.6~pc & \cite{Youngblood:2016ib} \\ 
$T_\mathrm{int}$ & Intrinsic temperature & $120$~K & \\
$T_\mathrm{irr}$ & Irradiation temperature & $790$~K & \\
$k_\mathrm{v}$ & Averaged opacity in the optical & $10^{-4.1}$~$\mathrm{cm^{2}}$~$\mathrm{g^{-1}}$ & \\
$k_\mathrm{th}$ & Averaged opacity in the infrared & $10^{-2.7}$~$\mathrm{cm^{2}}$~$\mathrm{g^{-1}}$ & \\ \hline
GJ~436b \\
$R_s$ & Host star radius & 0.464~$R_\odot$ & \citet{2008ApJ...677.1324T} \\
$M_p$ & Planet mass & $23.2$~$M_\oplus$ & \citet{2008ApJ...677.1324T} \\
$R_{1000}$~$_\mathrm{bar}$ & 1000-bar radius & 3.758~$R_\oplus$ & \\
$a$ & Semi-major axis & 0.02872~AU & \citet{2008ApJ...677.1324T} \\
$d$ & Distance & 10.1~pc & \cite{Youngblood:2016ib} \\ 
$T_\mathrm{int}$ & Intrinsic temperature & $170$~K & \\
$T_\mathrm{irr}$ & Irradiation temperature & $860$~K & \\
$k_\mathrm{v}$ & Averaged opacity in the optical & $10^{-3.6}$~$\mathrm{cm^{2}}$~$\mathrm{g^{-1}}$ & \\
$k_\mathrm{th}$ & Averaged opacity in the infrared & $10^{-2.3}$~$\mathrm{cm^{2}}$~$\mathrm{g^{-1}}$ & \\ \hline
HD~97658b \\
$R_s$ & Host star radius & 0.703~$R_\odot$ & \citet{2013ApJ...772L...2D} \\
$M_p$ & Planet mass & $7.86$~$M_\oplus$ & \citet{2013ApJ...772L...2D} \\
$R_{1000}$~$_\mathrm{bar}$ & 1000-bar radius & 1.943~$R_\oplus$ & \\
$a$ & Semi-major axis & 0.0796~AU & \citet{2013ApJ...772L...2D} \\
$d$ & Distance & 21.1~pc & \cite{Youngblood:2016ib} \\ 
$T_\mathrm{int}$ & Intrinsic temperature & $120$~K & \\
$T_\mathrm{irr}$ & Irradiation temperature & $1037$~K & \\
$k_\mathrm{v}$ & Averaged opacity in the optical & $10^{-4.1}$~$\mathrm{cm^{2}}$~$\mathrm{g^{-1}}$ & \\
$k_\mathrm{th}$ & Averaged opacity in the infrared & $10^{-2.7}$~$\mathrm{cm^{2}}$~$\mathrm{g^{-1}}$ & \\ \hline
Kepler-51b \\
$R_s$ & Host star radius & 0.940~$R_\odot$ & NASA Exoplanet Archive\tablenotemark{a} \\
$M_p$ & Planet mass & $2.1$~$M_\oplus$ & \citet{2014ApJ...783...53M} \\
$R_{1000}$~$_\mathrm{bar}$ & 1000-bar radius & 1.8~$R_\oplus$\tablenotemark{b} & \\
$a$ & Semi-major axis & 0.2514~AU & \citet{2014ApJ...783...53M} \\
$T_\mathrm{int}$ & Intrinsic temperature & $120$~K & \\
$T_\mathrm{irr}$ & Irradiation temperature & $793$~K & \\
$k_\mathrm{v}$ & Averaged opacity in the optical & $10^{-4.1}$~$\mathrm{cm^{2}}$~$\mathrm{g^{-1}}$ & \\
$k_\mathrm{th}$ & Averaged opacity in the infrared & $10^{-2.7}$~$\mathrm{cm^{2}}$~$\mathrm{g^{-1}}$ & \\
\enddata
\tablenotetext{a}{https://exoplanetarchive.ipac.caltech.edu}
\tablenotetext{b}{We use the value of 2.3~$R_\oplus$ for the clear atmosphere case}
\end{deluxetable*}


\section{Results} \label{results}
\begin{figure*}
\plotone{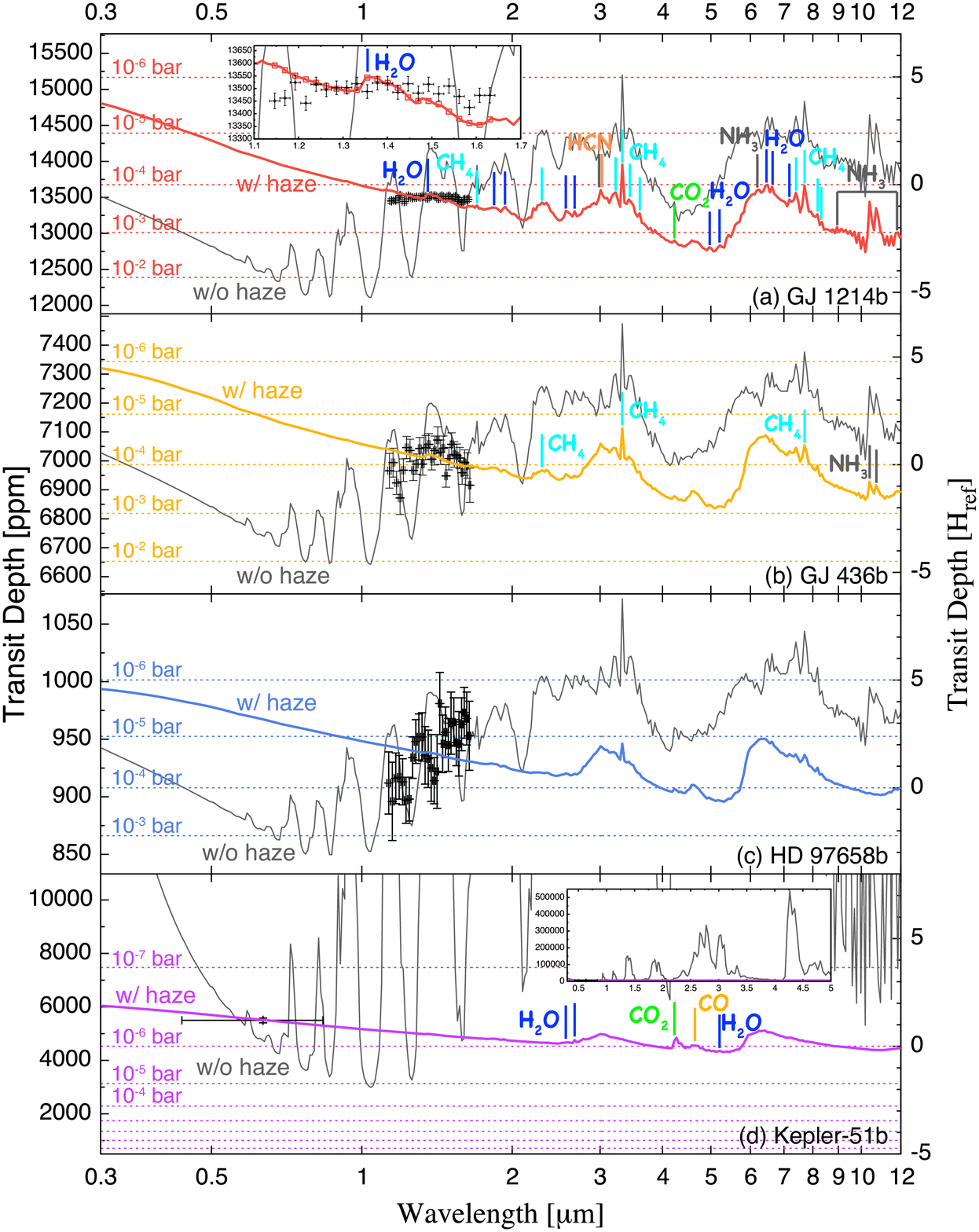}
\caption{Transmission spectrum models for atmospheres with haze (thick color lines) and without haze (thin gray lines) for the cases of GJ~1214b~(a), GJ~436b~(b), HD~97658b~(c), and Kepler-51b~(d). 
\ikoma{This is} in the order of the assumed total haze monomer production rate throughout the atmosphere, from the lowest \ikoma{panel} to the highest \ikoma{panel}, \ikoma{and} reflects the amount of UV flux each planet receives. 
\kawashimar{Those values are $1.38 \times 10^{-12}$, $7.83 \times 10^{-12}$, $5.45 \times 10^{-11}$, and $1.17 \times 10^{-10}$~g~$\mathrm{cm^{-2}}$~$\mathrm{s^{-1}}$ for GJ~1214b, GJ~436b, HD~97658b, and Kepler-51b, respectively.}
\kawashima{The left vertical axes show the transit depths in ppm, while the right ones show those in atmospheric scale height at a certain pressure level \kawashimar{for the hazy atmosphere cases} (see the text for details). 
Horizontal dotted lines represent the transit depths corresponding to the pressure levels of every one order for the hazy atmosphere cases \kawashimar{(see Eq.~(67) of \cite{2018ApJ...853....7K})}.}
Additionally, zoomed\ikoma{-}in and \ikoma{zoomed-}out view\ikoma{s} are shown for GJ~1214b~(a) and Kepler-51b~(d), respectively.
Observed transmission spectra of \cite{2014Natur.505...69K}, \cite{2014Natur.505...66K}, and \cite{2014ApJ...794..155K} for GJ~1214b, GJ~436b, HD~97658b, respectively, and the observed radius of Kepler-51b from \cite{2014ApJ...783...53M} are also plotted \kawashima{with black points} for reference.
\kawashima{Note that horizontal error bars are the bandpasses or the FWHM of those used.}
\kawashima{The red points in the zoomed-in view for GJ~1214b~(a) show the models binned at the observed bandpasses.} 
The transmission spectrum models are smoothed with the spectral resolution of $R = 100$.
\kawashimar{Note the broad bumps at 3.0, 4.6, and 6.3~{$\mathrm{\mu}$m} come from the absorption features of the tholin-like haze particles \citep{1984Icar...60..127K}.}
\label{fig-spectra}}
\end{figure*}

Figure~\ref{fig-spectra} shows the transmission spectrum models for atmospheres with and without haze for the cases of GJ~1214b, GJ~436b, HD~97658b, and Kepler-51b.
\kawashimar{We calculate the deviation in transit depth caused by $n H_\mathrm{ref}$ at a reference pressure level, $R_\mathrm{ref}$, as $2 n R_\mathrm{ref} H_\mathrm{ref} / R_s^2$ \citep{2001ApJ...553.1006B}, where $H_\mathrm{ref}$ is the atmospheric scale height at $R_\mathrm{ref}$.}
The reference pressures are taken to be $10^{-4}$~bar for GJ~1214b, GJ~436b, and HD~97658b, and $10^{-6}$~bar for Kepler-51b, which roughly correspond to the pressure-levels of the transit radii.
\kawashimar{Note that the transmission spectrum models for clear atmospheres are also calculated with the \kawa{distribution of gaseous species} 
from photochemical calculations.}

{
\kawa{A number of spectral features are produced by molecules above the optically-thick photochemical haze.}
These features are generally $\lesssim 2$ atmospheric scale height\kawa{s} for \kawa{GJ~1214b- and GJ~436b-like planets}, and $\lesssim 1$ for \kawa{HD~97658b-like planets}, \kawashimat{while $\lesssim 6$ for the cases of clear atmospheres}.} {As photodissociation is the driver for haze formation, this finding is general for exoplanets having \kawashimar{solar composition} atmospheres under similar levels of UV irradiation.


\kawashima{Considering a precision of $\sim$25~ppm expected to be achieved by JWST \citep{2014PASP..126.1134B} and a spectral resolution of $R = 100$, we find detectable absorption features with the upper-limit production rate of haze monomers for each planet as listed in Table~\ref{tab:features}.}

\begin{deluxetable*}{lll}
\tablecaption{Detectable absorption features by JWST\label{tab:features}}
\tablehead{
\colhead{Planet \kawa{Type}} &
\colhead{Molecules} &
\colhead{Wavelength [$\mathrm{\mu}$m]} \\
}
\startdata
GJ~1214b & $\mathrm{H_2O}$ & 1.4, 1.8, 1.9, 2.6, 2.7\kawashimar{, 5.0, 5.2, 6.5, 6.6, 7.2} \\
 & $\mathrm{CH_4}$ & 1.7, 2.3, 3.2, 3.3, 3.4, 3.6\kawashimar{, 7.4, 7.7, 8.2, 8.3} \\
 & $\mathrm{NH_3}$ & 3.0\kawashimar{, 6.2, 8.9-11.8} \\
 & HCN & 3.0 \\
 & $\mathrm{CO_2}$ & 4.2 \\ \hline
GJ~436b & $\mathrm{CH_4}$ & 2.3, 3.3\kawashimar{, 7.7} \\
 & \kawashimar{$\mathrm{NH_3}$} & \kawashimar{10.4, 10.7} \\ \hline
HD~97658b & - & - \\ \hline
Kepler-51b & $\mathrm{H_2O}$ & 2.6, 2.7\kawashimar{, 5.2} \\
 & $\mathrm{CO_2}$ & 4.3 \\
 & $\mathrm{CO}$ & 4.6
\enddata
\end{deluxetable*}


\kawa{Some molecular absorption features will be detectable for a planet like GJ~1214b}, due to lower incoming UV flux, namely lower monomer production rate}, and larger planet-to-star radius ratio.
{\kawa{For the planet GJ~1214b itself, however,} the 1.4~$\mathrm{\mu}$m $\mathrm{H_2O}$ feature was not detected by \cite{2014Natur.505...69K} at the precision of $\sim 25$~ppm.}
\kawashi{While \cite{2015ApJ...815..110M} demonstrated that their haze formation efficiency parameters of $\gtrsim 10$\% could match the observed flatness of \cite{2014Natur.505...69K} for 50 $\times $ Solar atmosphere, by considering particle growth, we have confirmed that our 1, 10, 100, 1000 $\times$ Solar atmospheres all fail to become as flat as the observation even adopting the maximum monomer production rates \citep{Kawashima.Ikoma2019}, possibly indicating extremely high metallicity, and/or \kawa{aggregate} haze particles \citep{2019ApJ...874...61A}, and/or coexistence of haze and other aerosols (see also discussion in \S\ref{discussion}).
\kawa{The message here is that a GJ~1214b-like planet having {solar composition} atmospheres {and similar UV environment} will be particularly suitable for future atmospheric characterization.}}
\kawashima{For the case of HD~97658b, all the features in the wavelength range 
are undetectable due to the higher incoming UV flux, namely higher monomer production rate, and smaller planet-to-star radius ratio.}

Among the detectable \kawashimar{molecules, it is promising that several features of $\mathrm{CH_4}$, which is a key indicator of haze formation, remains detectable even in the existence of haze for \kawa{GJ~1214b- and GJ~436b-like planets}, especially for its 3.3~$\mathrm{\mu}$m feature.}
\kawa{Note that this strong 3.3~$\mathrm{\mu}$m $\mathrm{CH_4}$ feature has been detected in the solar occultation spectrum of cooler celestial bodies in our solar system such as Saturn \citep{2015ApJ...814..154D} and Titan \citep{2014PNAS..111.9042R}.}
\ikoma{In the spectrum for Kepler-51b, the CH$_4$ features are invisible, because} \kawashima{$\mathrm{CH_4}$ is photodissociated and virtually absent above a thick haze layer at the low pressures of $\sim 10^{-6}$~bar}.

\ikoma{In} the case of Kepler-51b, the transmission spectrum model for the atmosphere with haze is quite featureless due to the existence of large haze particles in the upper atmosphere, 
unlike \ikoma{in} the other three cases.
\kawashima{However, distinct absorption features of $\mathrm{CO_2}$ and CO
\kawashimar{are detectable} \kawa{due to their larger abundances in the upper atmosphere}}.
\kawashimar{This} \kawa{comes from} its extremely low gravity, which \kawa{makes} an optical depth for a given pressure \kawa{much larger}
\kawa{and thus both the photodissociation rate of $\mathrm{CH_4}$ and resultant production flux of atomic carbon much smaller. 
This very small production flux of $\mathrm{C}$ allows effective conversion of C into $\mathrm{CO}$/$\mathrm{CO_2}$ by reaction between $\mathrm{H_2O}$, while $\mathrm{C}$ remains abundant and $\mathrm{CO}$/$\mathrm{CO_2}$ less abundant for the other planet cases.
}



\begin{figure*}
\gridline{
\fig{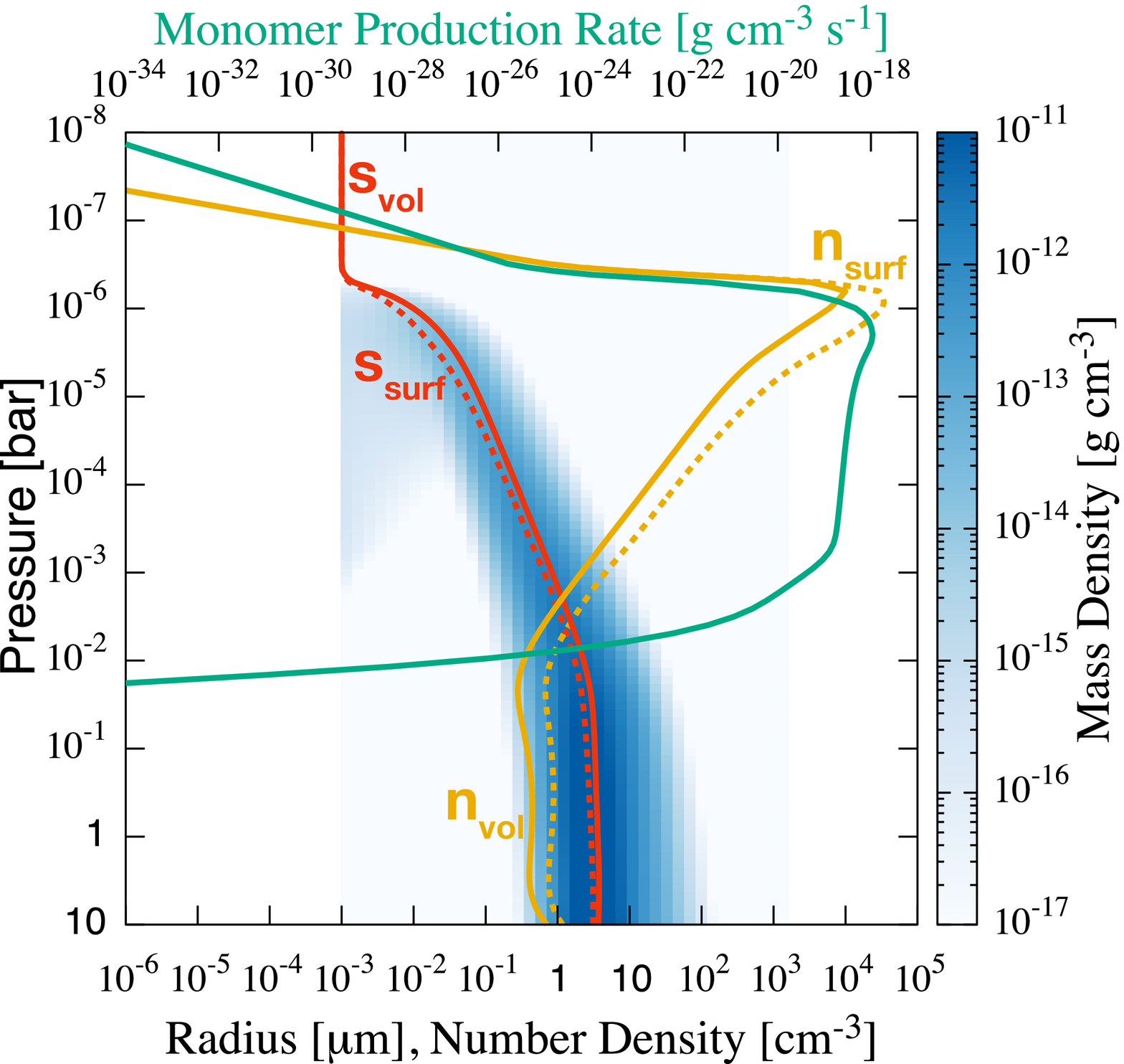}{0.45\textwidth}{(a)~HD~97658b \kawashima{($g = 1.1 \times 10^3$~$\mathrm{cm}/\mathrm{s}^2$ at $P = 10^{-6}$~bar)}}
\fig{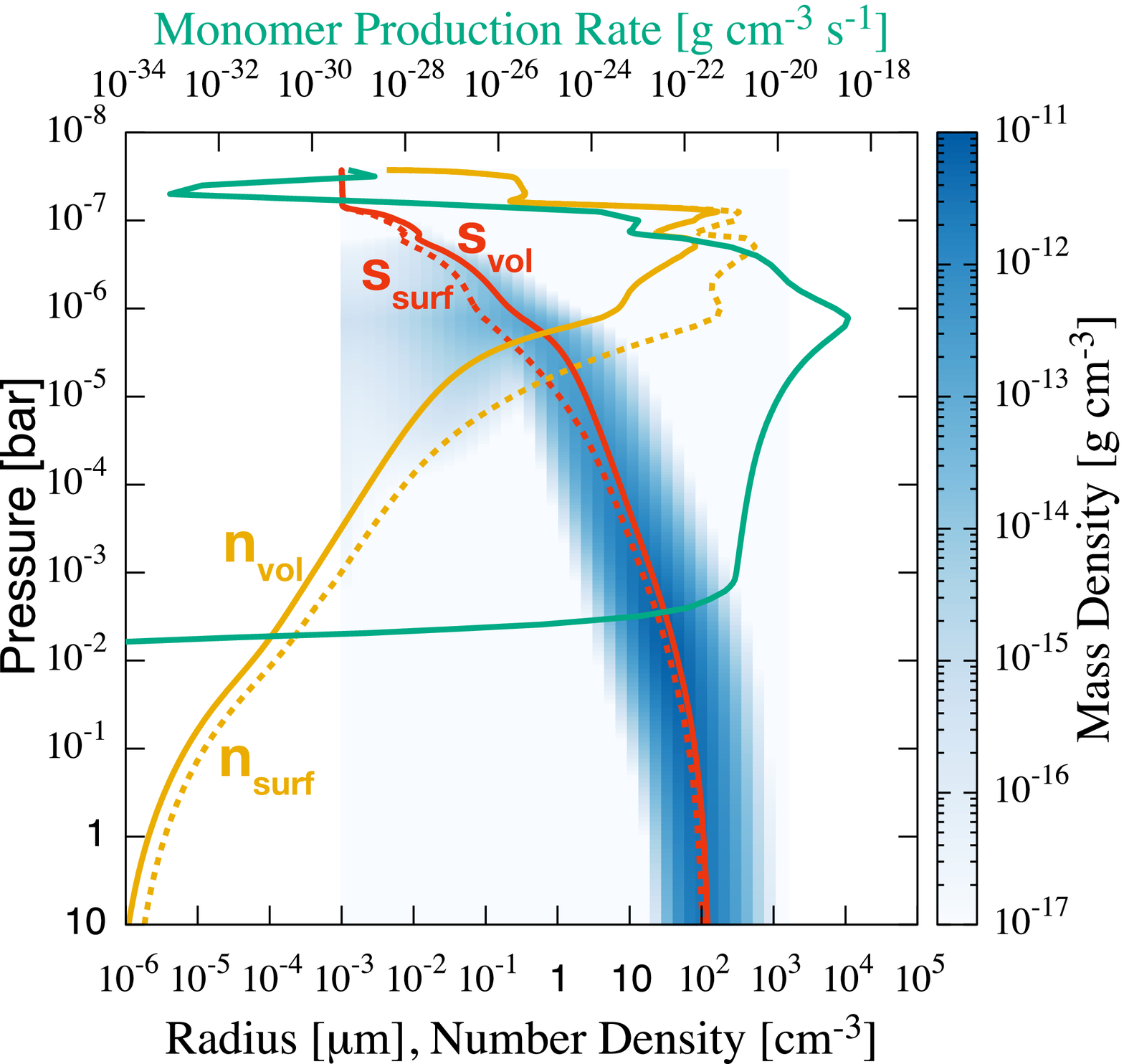}{0.45\textwidth}{(b)~Kepler-51b \kawashima{($g = 42$~$\mathrm{cm}/\mathrm{s}^2$ at $P = 10^{-6}$~bar)}}
}
\caption{Vertical profiles of \ikoma{the properties of haze particles including} the volume-average radius $s_\mathrm{vol}$ (red solid line) and number density $n_\mathrm{vol}$ (orange solid line), and the surface-average radius $s_\mathrm{surf}$ (red dashed line) and number density $n_\mathrm{surf}$ (orange dashed line), along with that of the monomer mass production rate (green solid line) for the cases of (a)~HD~97658b and (b)~Kepler-51b. See \cite{2018ApJ...853....7K} for the definition of each quantity.
The mass densities for all the size bins at each pressure level are also plotted with the blue color contour. 
\label{fig-growth}}
\end{figure*}

In Figure~\ref{fig-growth}, we plot the calculated vertical profiles of haze properties for the typical super-Earth gravity case of (a)~HD~97658b and the extremely low gravity case of (b)~Kepler-51b. 
In the case of HD~97658b, in the upper atmosphere \ikoma{($\lesssim$ 10$^{-6}$~bar)}, particles grow little because \ikoma{sedimentation occurs} faster than \kawashima{collisional} growth. 
\kawashima{Collisional} growth occurs significantly from the pressure level of $\sim 10^{-6}$~bar \ikoma{on} and results in \ikoma{haze particles with} the average radius of $\sim 3-4$~$\mu$m at the lower boundary of 10~bar. 
On the other hand, in the case of Kepler-51b, it is demonstrated that haze particles grow much larger in the upper atmosphere ($\sim 0.1$~$\mu$m at $\sim 10^{-6}$~bar) and results in the particles \ikoma{of} as large as $\sim 100$~$\mu$m at 10~bar. This is because particles can become large via \kawashima{collisional} growth even in the upper atmosphere\ikoma{,} instead of falling downward rapidly\ikoma{,} due to the small sedimentation velocity \renyu{from} low gravity.
Note that \ikoma{although the monomer production rate for Kepler-51b is $\sim 2$ times higher than that for HD~97658b, the effect of} the extremely low gravity \ikoma{turns out to make a much greater contribution to} \ikoma{such a} difference \ikoma{between those two spectra than that of high monomer production rate}. \kawashimar{Trends of vertical profiles for GJ~1214b and GJ~436b are similar to that of HD~97658b but with slightly smaller haze mass density because of the smaller monomer production rates \citep[see][for the dependence on monomer production rate]{2018ApJ...853....7K}.}


\section{Discussion} \label{discussion}


\begin{figure*}
\plotone{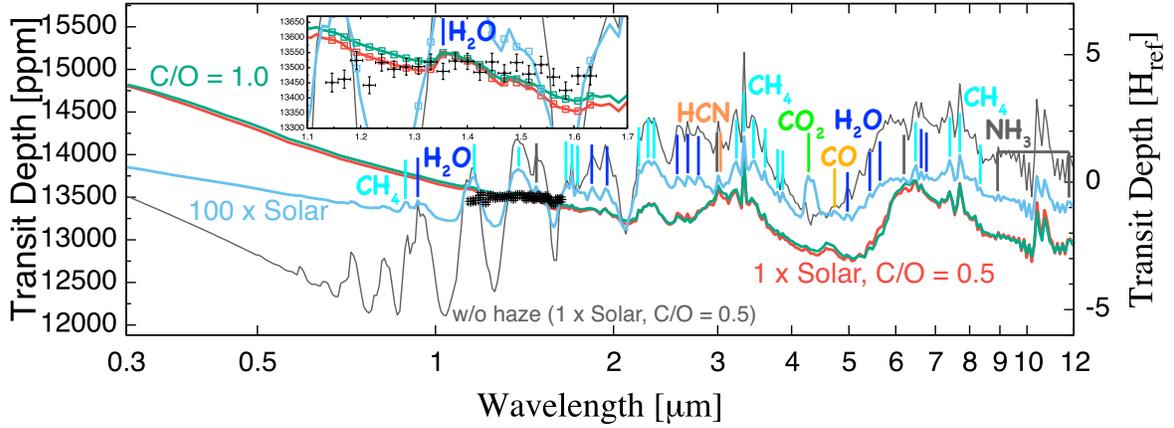}
\caption{\kawashimar{Same as Fig.~\ref{fig-spectra}~(a) but transmission spectrum models for the cases of 100 $\times$ Solar (light blue line) and $\mathrm{C/O = 1.0}$ (green line) atmospheres are also plotted. Detectable molecular absorption features for the case of 100 $\times$ Solar atmosphere are labeled.}
\label{fig-param}}
\end{figure*}

\renyu{\kawashimar{In this study, we have assumed maximum production rate of haze monomers in order to explore the detectable features at the worst case} \kawashimar{for solar composition atmospheres}}. 
\kawashimar{As for \kawa{the} eddy diffusion coefficient}, \kawashima{the molecular absorption features become more prominent for the larger value because of the efficient downward transport of haze particles, while \kawa{remaining} similar for the smaller value \citep{Kawashima.Ikoma2019}.}

\kawashimar{In order to briefly assess the effect of metallicity and C/O ratio, in Figure~\ref{fig-param}, we show the transmission spectrum models for the cases of different metallicity and C/O ratio for GJ~1214b. One realizes that more absorption features become detectable for the case of 100 $\times$ Solar atmosphere because of the smaller monomer production rate due to an enhanced photon-shielding effect by other molecules, while some features at relatively long wavelengths become undetectable due to smaller atmospheric scale height \citep{Kawashima.Ikoma2019}. On the other hand, \kawa{the} effect of \kawa{the} C/O ratio is relatively small and among the detectable features for the solar composition atmosphere, $\mathrm{H_2O}$ features at 1.8, 5.0, 5.2, and 7.2~$\mu$m, $\mathrm{CO_2}$ features at 4.2~$\mu$m, and $\mathrm{NH_3}$ features at 6.2~$\mu$m become undetectable \kawashimar{due to the slightly higher monomer production rate} \citep{Kawashima.Ikoma2019}.}

Besides hydrocarbon haze, the existence of other aerosols can {mute spectral features in transmission}: condensation clouds such as \kawashimar{KCl, ZnS, $\mathrm{K_2SO_4}$, ZnO, and graphite clouds} in the temperature range of interest in this study ($\lesssim 1000$~K) \kawashimar{\citep{2012ApJ...745....3M, 2013ApJ...775...33M, 2016ApJ...827..121M, 2018ApJ...859...34O, 2018ApJ...863..165G, 2019A&A...622A.121O}} and photochemical haze made of sulfur species \citep{2013ApJ...769....6H, 2016ApJ...824..137Z, 2017AJ....153..139G}.


\kawashima{In outflowing atmospheres, while having little effect on 
\ikomaf{the profiles of pressure and density}, the velocity of the outward flow may affect the sedimentation of small particles.
The sedimentation velocity of haze particles are much faster than the outward velocity, which we estimate with the isothermal Parker solution \citep{1958ApJ...128..664P} for simplicity, for all the cases except Kepler-51b, \ikomaf{whereas} we have confirmed that {in the case of Kepler-51b,} the sedimentation velocity of the volume-averaged-size particles are smaller than the outward velocity for the region of the pressures lower than $1.3 \times 10^{-7}$~bar. However, since the dominant monomer production region is below this pressure-level, 
\ikomaf{the flow} would have little impact on our results. The detailed consideration of this effect is beyond the scope of this study.}

\kawashima{Also, \ikomaf{as} for 
extended atmosphere\ikomaf{s} like Kepler-51b\ikomaf{'s}, the effect of the tidal potential may affect the hydrostatic structure of the atmosphere. By using Eq.~(12) of \cite{2007A&A...472..329E}, we have confirmed that the gravitational potential at the optically thick radius at $1.0 \times 10^{-6}$~bar, which is 9\% of its Hill radius, would decrease to 86\% due to the tidal potential and thus do not have significant impact on our results.}

\section{Conclusions} \label{conclusions}
In this study, we have used four warm ($\lesssim 1000$~K) planets suitable for atmospheric characterization via transmission spectroscopy, GJ~1214b, GJ~436b, HD~97658b, and Kepler-51b, as examples and explored molecular absorption features detectable by JWST even in the existence of hydrocarbon haze in their atmospheres. 
\ikoma{Using the 
\kawa{models of} \citet{2018ApJ...853....7K},} we have simulated photochemistry, the growth of hydrocarbon haze particles, and transmission spectra for the atmospheres of these four planets.

We have found
\kawa{that among the planetary parameters considered, {super-Earths with hazy, relatively hydrogen-rich atmospheres} are mostly expected to produce detectable molecular absorption features {such as a quite prominent $\mathrm{CH_4}$ feature at 3.3~${\rm \mu}$m}} \kawa{even for the {extreme} case of the {most efficient} production of photochemical haze.}
{The sizes of those features correspond to $1\sim2$ atmospheric scale heights; while they are substantially smaller than a haze-free atmosphere, those features would be detectable with at precision expected for JWST.}
\kawashimar{Especially, planets with higher gravity, lower UV irradiation, and higher temperature are \kawa{more} suitable \citep{Kawashima.Ikoma2019}.}
\kawa{{Slight disagreement between our synthetic spectrum and very flat one observed for GJ~1214b, however,} implies the importance of additional confounding effects.} 

We have also demonstrated that in the case of extremely low gravity planet, Kepler-51b, haze particles grow significantly large in the upper atmosphere due to the small sedimentation velocity, resulting in the featureless or flat transmission spectrum \kawa{in a wide wavelength range}.

In summary, various molecular absorption features are \kawa{expected to be} detectable \kawa{for relatively hydrogen-rich atmospheres} even in the existence of hydrocarbon haze 
\ikoma{with} JWST\ikoma{, given its high precision} \renyu{and long-wavelength capabilities.}
\renyu{We thus suggest that the transmission spectra with muted features measured by 
\kawa{HST} in most cases do not preclude strong features at the longer wavelengths accessible by JWST.}

\acknowledgments
\kawa{We thank the anonymous referee for his/her careful reading and constructive comments, which helped us improve this letter greatly.}
\renyu{The research was carried out at the Jet Propulsion Laboratory, California Institute of Technology, under a contract with the National Aeronautics and Space Administration.}
Y.K. thanks the travel support of JPL's Science Visitor and Colloquium Program.
Y.K. is supported by the Grant-in-Aid for JSPS Fellow (JSPS KAKENHI No.15J08463), Leading Graduate Course for Frontiers of Mathematical Sciences and Physics, Grant-in-Aid for Scientific Research (A) (JSPS KAKENHI No.15H02065), and the European Union's Horizon 2020 Research and Innovation Programme under Grant Agreement 776403.
\kawa{R.H. is supported by NASA Grant No. 80NM0018F0612.}
\kawa{M.I. is supported by JSPS KAKENHI No.~18H05439 and Core-to-Core Program International Network of Planetary Sciences.}
This work has made use of the MUSCLES Treasury Survey High-Level Science Products (doi:10.17909/T9DG6F) and the SVO Filter Profile Service (http://svo2.cab.inta-csic.es/theory/fps/) supported from the Spanish MINECO through grant AyA2014-55216.

\end{document}